\documentclass[aps,pra,groupedaddress,showpacs,preprint]{revtex4}
\usepackage{graphicx}

\newcommand{\be}{\begin{equation}}
\newcommand{\ee}{\end{equation}}
\newcommand{\bea}{\begin{eqnarray}}
\newcommand{\eea}{\end{eqnarray}}
\newcommand{\beb}{\begin{eqnarray*}}
\newcommand{\eeb}{\end{eqnarray*}}

\begin{document}

\title{Efficient Random Walk  Algorithm for Simulating Thermal Transport in Composites With High Conductivity Contrast}
\author{K.G.S.H. Gunawardana}
\author{Kieran Mullen}
\email[Electronic address: ]{mullen@mail.nhn.ou.edu}
\affiliation{Homer L. Dodge Department of Physics and Astronomy, The University of
Oklahoma, 440 West Brooks Street, Norman, Oklahoma 73019-0225}
\author{Dimitrios V. Papavassiliou}
\affiliation{School of  Chemical,Biological and Material  Engineering, The University of Oklahoma, 100 East Boyd
Street, Norman, Oklahoma 73019-1004}

\begin{abstract}
In dealing with thermal transport in composite systems, high contrast materials pose
a special problem for numerical simulation: the time scale or step size in
the high conductivity material must be much smaller than in the low
conductivity material. In the limit that the higher conductivity inclusion
can be treated as having an infinite conductivity, we show how a standard
random walk algorithm can be alterred to improve speed while still
preserving the second law of thermodynamics. We demonstrate the principle in
a 1D system, and then apply it to 3D composites with spherical inclusions.

\end{abstract}

\maketitle

%
% =============================================================================

% =============================================================================

% =============================================================================

\section{Introduction \label{sec:Introduction}}

A variety of systems display dramatically different thermal 
diffusivities.  For example, 
the thermal conductivity is estimated at 3000 W/mK for an isolated
multiwall carbon nanotube (CNT) and between 1750 and 6600 W/mK for a
single wall carbon nanotube at room temperature.\cite{ESChoi, JChe,
JHone,SBerber}   Typical polymer matrices,
 in contrast, have thermal conductivities that are three orders of
magnitude smaller.  Composites using carbon nanotubes have been
suggested as cheap materials with average thermal
conductivity.\cite{RBBird} However making improved thermal conducting
 polymer composites has been hindered due to the Kapitza thermal
 resistance and various processing issues.  

It may be possible to
alter this resistance through functionalizing the ends or surface
of the CNTs, although this may decrease the thermal conductivity of
the material.  
The problem of optimizing  the thermal conductivity of
CNT composites presents an intriguing combination of high conductivity
contrast, strong disorder, 
and incorporated materials with an extremely 
high aspect ratio.
Thus an efficient and reliable method to calculate effective thermal
 conductivity and varying interface resistance
 in this two phase medium is desirable.

This problem has been studied in the context of electrical
conductivity.\cite{elect1,elect2}
However, the inclusions in thermal transport can have be
quite asymmetric and entangled.   In these cases it the isotropic
averaging models may not apply.

One approach to model the thermal transport in composites is to use
a random walk algorithm in which the transport is assumed to be
diffusive.  For instance, Tomadakis and Sotirchos \cite{Toma1,Toma2}
has used this approach to find the effective transport properties
of random arrays of cylinders
 in a conductive matrix. Recently, Doung et.al \cite{cnt1, cnt2,cnt3}
 developed a random walk algorithm to model  thermal transport in
 carbon nanotube-polymer composites and the simulation results
 showed a reasonable agreement with the experimental data for
 Epoxy-SWNT composites\cite{cnt2}.  In this approach thermal transport
 is described by random jumps of thermal markers carrying a certain
 amount of energy ($\Delta E$). The step size($\Delta x=x_2-x_1$) of this
 thermal markers follows the gaussian distribution.(See Eq.\ref{eq:oneDgaussian})
The standard deviation ($\sigma $) of the gaussian step distribution
in each one of the space dimensions is $\sigma_{M/I}=\sqrt{2 D_{M/I}
\Delta t}$, where $\Delta t$ is the time increment, $M/I$  refers
to matrix or inclusions and $D_{M/I}$ is the thermal diffusivity.
However problem arises when their is a high contrast in thermal
diffusivity of matrix and inclusions. The step size  in highly
conducting inclusions $\Delta x_{I}$ become very large compare to
the that of poorly conducting matrix  $\Delta x_{M}$. Eventually
this leads the markers to jump out the inclusions as soon as they
enter.  This can be avoided having very small steps  inside the
matrix so that the steps inside the inclusions are within the dimensions
of the inclusions. But this is  computationally expensive. 

When the thermal diffusivity of the inclusion is very high relative
to the matrix it is
reasonable to assume that the thermal diffusivity of the inclusions is 
infinite.  This obviates the
need to model random walks inside the inclusions. In this approach
markers entering an infinite conductivity inclusion (ICI)
are distributed uniformly inside the
inclusion on the next time step.  Some fraction will leave on the next
time step and they 
always leave from the surface of the inclusion.  (Otherwise the simulation
wastes time on walkers that hop within the ICI.)
However, we must be careful in choosing how the walkers leave the
the ICI since incorrect approaches can lead
to the unphysical result of a system at uniform temperature
spontaneously developing a temperature gradient at the interface
between the inclusion
 and the medium.\cite{cnt2} While the effect is apparently small,
 it must
be remembered that diffusion occurs at these same interfaces. In
this paper we provide a rigorous approach for implementing a random
walk algorithm with emphasis on the treatment at the interface
between the inclusions and the matrix material for high conductivity
contrast composites, and we quantify the errors made when gaussian
and modified step distributions are employed.

This paper is divided into four parts. In the first, we briefly
describe  the algorithm for ``infinite conductivity''  inclusions.
Next, we show the rigorous way to handle inclusions in one dimensional
systems. We verify our results numerically in ordered and disordered
systems, and compare them to results obtained by assuming that the
walkers leave the surface with a gaussian step distribution. In the
next section we develop this approach to spheres in three dimensions
and again verify it numerically, showing quantitatively the errors
that develop if a gaussian step distribution from the surface is used. 
Interestingly,
the errors in thermal conductivity are larger in 3D than in 1D and
larger for random arrays than for regular ones. In the final section
we conclude with a summary and a discussion of future work.

\section{The Model}

The goal is to calculate the thermal conductivity of a composite
composed of a matrix containing a distribution of 
``infinite conductors'' (ICs).  This conductivity is calculated by
fixing the   heat flux through the computational volume and measuring
the resulting average temperature gradient.
The diffusion of heat is modelled by the motion of random walkers
within the domain.
The computational cell is divided in bins, and the temperature
distribution is calculated from the number of walkers in each bin.
To maintain a constant heat flux in the $x$ direction through the computational cell, 
random  walkers  carrying $+\Delta E$ energy are periodically 
added  at the surface $x=x_{\mathrm{min}}$, and then allowed to
move with random jumps that follow a gaussian distribution into the
computational cell.  In order to fix an ``outward'' energy flux on the
opposite surface, 
random walkers carrying  $-\Delta E$  energy are added  at the
surface $x=x_{\mathrm{max}}$  at the same rate as the  positive markers.
The   $+\Delta E$ and $-\Delta E$ thermal markers are often called
"hot" and "cold" walkers. The exact size of $\Delta E$ is arbitrary: the
heat flux might be modelled by many small walkers or one large one.
However using too many walkers is computionally ineffcient, while too few
produces noisy results that requires more runs to get better averages.
In the  $y$ and $z$ direction the
computaional domain is assumed to be periodic.  The solution at steady
states yields a linear temperature profiile and the  thermal
conductivity can be extracted from Fourier's law.  To incorporate
the effect of the Kapitza thermal resistance, walkers 
in the matrix that would normally attempt to jump into the IC 
can only do so with a probability $f_{m,IC}$.\cite{Kap,Shen,JLB,CJTwu} 
Thus they stay in the  matrix phase with a probability $1-f_{m,IC}$.
 The value of $f_{m,IC}$ is determined by the Kapitza resistance.
This can be estimated using
acoustic mismatch model when the physical properties of the
materials are known.\cite{SchwartzPohl}

Similarly, random walkers located within the IC have a probability to
hop out on each time step. 
Exactly what fraction of the walkers
should leave in each time step, and the exact nature of the probability
distribution for the steps they should take from the surafce are determined
 in the next two sections.  However, those that do leave, exit at random
positions on the IC.  This is done to model the ``infinite'' conductivity
of  IC so that the walker distribution within the IC is uniform. 

Collisions between walkers are ignored. The random walk reflects the
scattering of phonons in the disordered matrix material. Walker-walker
scattering would reflect nonlinear thermal conductivities which are
typically small.
Similarly, we assume that the 
properties of the materials (e.g. density, specific heat, thermal
relaxation length) do not change with temperature over the range modelled.

Finally, we assume that 
the product of the mass density of IC and specific heat capacity equals that of the
matrix, so that in thermal equilibrium the walker density would be uniform
inside and out of the IC's. This is done for simplicity, so that the local
temperature is simply proportional to the difference of the average density
of hot and cold walkers.  Without this
assumption we would have to alter the probability of walkers entering and
leaving the IC's so that in thermal equilibrium, the ratio of average walker
density inside the IC to that of the matrix equals the ratio of their 
volumetric heat capacities.  Only then would the equilibrium walker
distribution represent a uniform temperature.

\section{Random walks with infinite conductivity inclusions in 1D. 
\label{sec:1D}}

Below we describe how to efficiently handle the random walks in a fashion
that satisfies the second law of thermodynamics. The difficulty lies in
properly handling the random walkers that jump out from the high
conductivity material. To make the explanation clear, we first look at the
one dimensional case. We subsequently address the three dimensional case for
spherical inclusions in section \ref{sec:3D}.

\subsection{Analytic results for one dimensional walks \label{subsec:oneD}}

\begin{figure}[bt]
\centering
\includegraphics[scale=0.5]{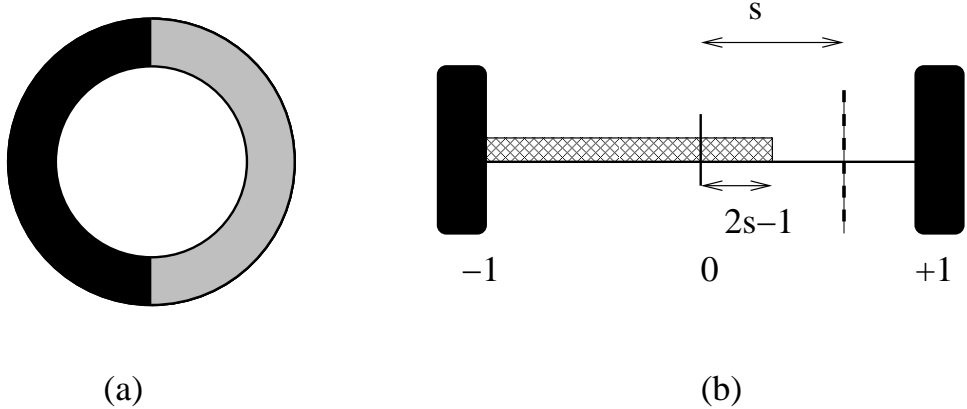}
\caption{{\protect\small An illustration of the one dimensional model. In
(a) the darker material on the left is of ``infinite'' conductivity. The one
dimesional system therefore has two boundaries as shown in (b), where we
assume $\protect\sigma<<1$. The difficulty is that in equilibrium the net
flux through any surface (e.g. the dashed line) must be zero. Walkers
hopping through the surface from the shaded region in  Fig.(b) must be
balanced by walkers leaving the right hand infinite conductor. This implies
that if random walkers always leave the \textit{surface} of the infinite
conductor (IC), they must have a different jump distribution than random
walkers inside the ``normal'' region. }}
\label{fig:oneDmodel}
\end{figure}

We consider a set of random walkers moving in a one dimensional ring, half
made from an ``infinitely conducting material'' as show in 
fig.\ref{fig:oneDmodel}. We can view this as 
a one dimensional line with boundaries
at $x=\pm 1$. We know that in equilibrium the density of random walkers
throughout the whole ring should be uniform.

Consider a surface located at $x=s$, as shown in fig.\ref{fig:oneDmodel}.b.
The flux of random walkers from the left through the surface must equal that
from the right. This is not a problem for a surface located near the center
of the interval. However, if $1>\sigma >1-s$, the flux of random walkers from right
 \textit{in the matrix medium} cannot balance those from the
left; there are too few of them. The solution lies in that the difference
must be made up from random walkers leaving the \textquotedblleft
infinite\textquotedblright\ conductivity material. If they were distributed
uniformly throughout the infinite conducting material, their flux would
maintain the equilibrium.

We do not wish to model the inside of the IC inclusions because random
walkers within them move on a much faster time scale than those outside. We
assume that a random walker instantly leaves from any point on the surface
(in this case from $x=\pm 1$). However, since they leave always exactly from
the surface, their step distribution must be different from that of random
walkers within the matrix medium.

In each step of the simulation we move the walkers inside the interval $
-1<x<1$, as well as those outside. We wish to do this in a fashion that is
in agreement with the second law of thermodynamics. Let the probability that
a walker in the matrix medium jumps from $x_{1}$ to $x_{2}$ be given by: 
\begin{equation}
P(x_{1},x_{2})={\frac{1}{\sqrt{2\pi }\sigma }}e^{\frac{-(x_{2}-x_{1})^{2}}{
2\sigma ^{2}}}  \label{eq:oneDgaussian}
\end{equation}
In each time interval we can see that only a fraction of the walkers inside
the IC will leave, or else their density would not equal that in the normal
medium. In this simple model, we require that the number inside the matrix
medium ($N_{i}$) equals that outside ($N_{o}$) in the IC. We also require
that the net flux through a surface located at $x=s$ is zero.

\vspace{25pt} The flux to the left from particles lying in the matrix
region, $s<x<1$ is balanced by the flux to the right for those lying between 
$2s-1<x<s$. However those in the shaded region of fig.\ref{fig:oneDmodel}
are not so compensated. They must be balanced by a net flux of walkers
leaving the righthand boundary. Denote the flux from the shaded region to
the right by $N_{r}$; it is : 
\begin{eqnarray}
N_{r} &=&\rho _{0}\int_{-1}^{2s-1}dx_{1}\int_{s}^{\infty
}dx_{2}P(x_{1},x_{2}) \\
&=&{\frac{\rho _{0}}{2}}\int_{0}^{\infty }\mathrm{\,Erfc\,{\left( \frac{z-s+1
}{\sqrt{2}\sigma }\right) }}\,dz
\end{eqnarray}
where $\mathrm{\,Erfc\,{(x)}}$ is the complementary error function, $\mathrm{
\,Erfc\,{(x)}}=1-\mathrm{\,Erf\,{(x)}}$.
We are summing over any walker starting in the shaded region ending up
anywhere to the right of the barrier. We
have let the upper limit of the endpoint of the jump to infinity since $
\sigma <<1$; we extend the lower limit of the first integral to $-\infty $,
and we have shifted variables to $z=2s-1-x_{2}$. We neglect any walkers
leaping from the IC on the left boundary $x=-1$ all the way through $s$.

The flux $N_{r}$ must be balanced by the flux from walkers leaving the IC on
the right. Let the probability that a random walker in the IC leaves it be
given by $\lambda $, and the probability that it jumps to a point $x$,
leaving from the right hand boundary, be $f(x)$. Then the flux to the left
through the surface at $x=s$ due to these walkers is 
\begin{equation}
N_{\ell }=N_{o}\lambda \int_{-\infty }^{s}f(x)\,dx
\end{equation}
We set $N_{\ell }=N_{r}$, and take the derivative of both sides with respect
to $s$. This gives us an integral expression for $f(s)$: 
\begin{equation}
f(s)={\frac{\rho _{0}}{2N_{0}\lambda }}\left[ \mathrm{\,Erf\,{\left( \frac{
s-1}{\sqrt{2}\sigma }\right) }}+1\right] 
\end{equation}
The requirements that $N_{i}=N_{o}$ and the balancing of the fluxes when $s=1
$ is enough to solve for $f(s)$. The distribution of steps, $\tilde{f}
(u)\equiv f(1-u)$ is given by: 
\begin{equation}
\tilde{f}(u)=\sqrt{\frac{\pi }{2}}{\frac{1}{\sigma }}\left( 1-\mathrm{\,Erf\,
{\left( \frac{u}{\sqrt{2}\sigma }\right) }}\right)   \label{eq:oneDstep}
\end{equation}

\subsection{Numerical results for thermal conductivity in 1D \label{subsec:1Dtc}}

\begin{figure}[bt]
\centering
\includegraphics[scale=0.35]{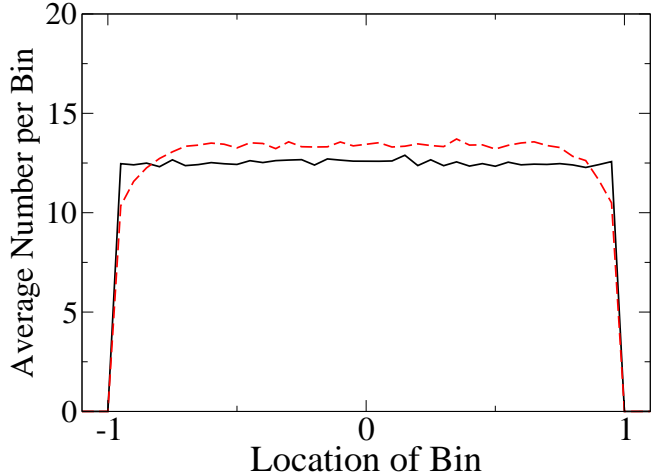}
\caption{{\protect\small (Color online) Results for the one dimensional model of a random
walk on a ring with an infinite conductivity inclusion. The steps in the
random walk have a Gaussian distribution with a mean of $0.20$. Plotted are
the average number of walkers in each of 20 bins, equally spaced $-1<x<1$
over the course of $10^5$ Monte Carlo steps. Starting with 500 random
walkers, half should be in the ``normal'' region, so that the average
number/bin should be 12.5. The dashed line is the result of performing the
simulation incorrectly, and letting the walkers have a Gaussian step
distribution as in eq.\protect\ref{eq:oneDgaussian}; There are too many
walkers in the interval, and their distribution is not uniform. The solid
line is the result of using eq.\protect\ref{eq:oneDstep}, which yields the
correct result. }}
\label{fig:oneDres}
\end{figure}

The above analytical calculation provides the correct step distribution for
walkers leaving the edge of the infinite conductors. We can compare it to a
simple model where we simply have the walkers take a step with a Gaussian
probability distribution (mean size 0.20) from the surface. Fig.\ref{fig:oneDres} is the spatial distribution of random walkers in such a one
dimensional system.\cite{probability} Plotted are the average number of
walkers in each of 20 bins, equally spaced $-1<x<1$ over the course of $
10^{5}$ Monte Carlo steps. Starting with 500 random walkers, half should be
in the \textquotedblleft matrix\textquotedblright\ region, so that the
average number/bin should be 12.5. The dashed line is the result of
performing the simulation incorrectly, and letting the walkers have a
Gaussian step distribution as in eq.\ref{eq:oneDgaussian}; There are too
many walkers in the interval, and their distribution is not uniform. The
solid line is the result of using eq.\ref{eq:oneDstep}, which yields the
correct result, and is uniform.

\begin{figure}[bt]
\centering
\vbox{\includegraphics[width=3.25in]{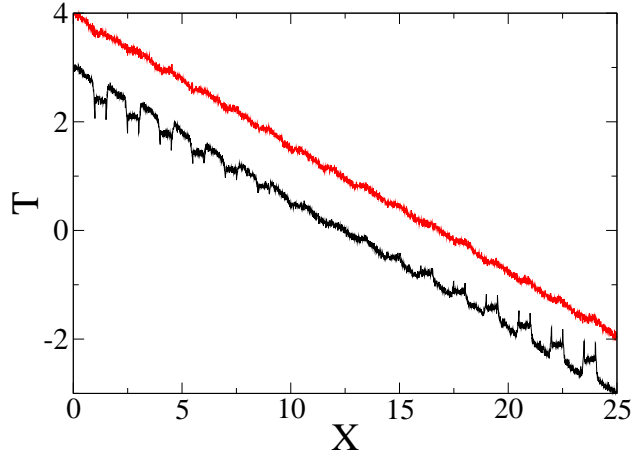}}
\caption{{\protect\small (Color online) Plots of the temperature distribution in a periodic
array of infinite conductivity inclusions in a 1D matrix. The temperature is
determined by the average of the difference between the number of positive
and negative walkers in a given region. This plot was generated over a run
of 62,500 time steps where 10,000 positive (negative) walkers were added
every 10 time steps at the left (right) border. The ICs are 0.50 units long
and the matrix spacer between them is 1.00 units. The probability to cross
into an IC from the matrix is 0.16. Lower line: The temperature distribution
generated by having random walkers depart the ICs with Gaussian steps. Upper
line: The temperature distribution (shifted up one unit for clarity) given
by an algorithm using the step probability distribution of 
eq.\protect\ref{eq:oneDstep}. Note that temperature gradients spontaneously appear at
interfaces when the incorrect jump distribution is used. }}
\label{fig:1Ddens}
\end{figure}

In order to determine the significance of this error, we place several ICs
in the computational volume and run at constant heat flux until the
temperature distribution converges. We then extract the gradient in walker
density and calculate the thermal conductivity. Sample results are plotted
in fig.(\ref{fig:1Ddens}), where we show the results for Gaussian steps
(lower curve) and steps governed by eq.(\ref{eq:oneDstep}) (upper curve).
The latter gives physically reasonable results (with noise), in which the
temperature is constant the ICs and uniformly decreasing in the matrix.

\begin{figure}[bt]
\centering
\vbox{\includegraphics[width=3.25in]{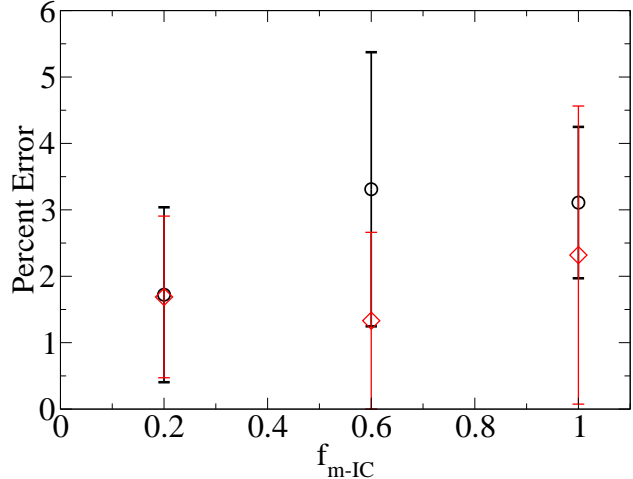}}
\caption{{\protect\small (Color online) Plot of the relative percent error for the the
thermal conductivity calculated using a Gaussian step distribution ($\protect\sigma=0.1$) as compared to that of eq.(\protect\ref{eq:oneDstep}) as a
function of $f_{m-IC}$, the probability for a walker to enter into an
inclusion. The results are for a one dimensional system with 20 inclusions;
the diamonds are for a regular array of ICs and the circles are random ICs.
The error bars are based on a sample of five different configurations and
are included to give an indication of how large the the errors can be. }}
\label{fig:1Dvar}
\end{figure}

The thermal conductivity is extracted from the ratio of the slope of the
temperature (the walker density) to the applied flux. In fig.(\ref{fig:1Dvar}
) we plot the average value of the percent error in the thermal conductivity
as a function of the transmission probability, $f_{m,IC}$, for regular and
random 1D arrays. In this simulation the ICs were 0.50 units long and the
material between them was 1.00 units wide.
The results are averaged over five runs
each lasting for 40,000 time steps. The percent error is defined as the
difference between the results of simulations using the Gaussian steps and
the results using eq.\ref{eq:oneDstep}. The error bars represent the
variation in thermal conductivities over the runs. Thus we see that the
error can range as large as five percent, and that can vary substantially.

\section{Three dimensional models \label{sec:3D}}

We have shown above that errors in one dimensional simulations are
avoidable, but only a few percent. Below we generalize the above problem to
three dimensions for spherical inclusions, and show that the effect can be
significant.

\subsection{Analytic Derivation of Random Walks with Spherical Inclusions 
\label{subsec:Sphere}}

In our model random walkers that land inside the sphere are immediately
moved to a random point on the surface of the sphere. On the next time step
they can move in the radial direction away from the sphere. We assume that
if we choose the fraction that leave and their step distribution correctly,
then when we are in equilibrium we will obtain a uniform, stationary density
outside and inside the sphere.

The number of random walkers entering the sphere from a region $d\vec{r}$
near $\vec{r}\,$ landing inside the sphere is given by: 
\begin{equation}
n(\vec{r})=\,d\vec{r}\,\,\,{\frac{\rho _{0}f_{k}}{(2\pi )^{3/2}\sigma ^{3}}}
\int_{r^{\prime }<R}d\vec{r^{\prime}}\,e^{\frac{-(\vec{r}-\vec{r^{\prime}})^2}{2\sigma^2}}
\end{equation}

where we have generalized the probability distribution of eq.(\ref{eq:oneDgaussian}) to three dimensions. The factor $f_{k}$ represents the
Kapitsa resistance; it is the probability that a random walker will enter
the spherical inclusion. When $f_{k}=0$, no walkers enter the inclusion and
the Kapitsa resistance is infinite; when $f_{k}=1$ then walkers can freely
step into the inclusion and the Kapitsa resistance is zero. The total number
entering a sphere of radius $R$ is 
\begin{equation}
N_{\mathrm{in}}(R)=\int_{r>R} d\vec{r}\,\,\,{\frac{\rho _{0}f_{k}}{(2\pi )^{3/2}\sigma ^{3}}}
\int_{r^{\prime }<R}d\vec{r^{\prime}}\,e^{\frac{-(\vec{r}-\vec{r^{\prime}})^2}{2\sigma^2}}
\end{equation}
For any value of ${\vec{r}}$ we can rotate our primed coordinate system so
that $\hat{z}^{\prime }\parallel {\vec{r}}$ so that the angle between ${\vec{
r}}$ and ${\vec{r}\,^{\prime }}$ is simply $\theta ^{\prime }$, the
spherical polar angle in the primed system. The angular integrals can then
all be done in closed form giving 
\begin{equation}
N_{\mathrm{in}}(R)=\sqrt{8\pi }\,{\frac{\rho _{0}f_{k}}{\sigma }}
\int_{r>R}r\,dr\int_{r^{\prime }<R}r^{\prime }dr^{\prime }\,\left( e^{\frac{
-(r+r^{\prime})^2}{2\sigma ^{2}}}-e^{\frac{-(r-r^{\prime})^2}{2\sigma ^{2}}
}\right) 
\end{equation} 
The resulting integral can also be found exactly yielding 
\begin{equation}
N_{\mathrm{in}}(R)={\frac{2}{3}}\rho _{0}f_{k}\left[ \sqrt{2\pi }\sigma
\left( 3R^{2}-\sigma ^{2}\right) +2\pi R^{3}\,\mathrm{\,Erfc\,{({\frac{\sqrt{
2}R}{\sigma }})}}+\sqrt{2\pi }\sigma e^{\frac{-2R^{2}}{{\sigma ^{2}}}
}(\sigma ^{2}-R^{2})\right]   \label{eq:Nin}
\end{equation}

If the density of walkers is uniform then the number inside the sphere is $
V_s \rho_0$ where $V_s$ is the volume of the sphere.
(This is only true when the product of the density and specific heat capacity of the matrix and IC's are equal. 
When this constraint does not hold the number of walkers inside the IC is 
$V_s \rho_0 \frac{C_M \rho_M}{C_{IC} \rho_{IC}}$. Where $C_{IC}$($C_{M}$) and $\rho_{IC}$($\rho_{M}$) are specific heat capacity and mass  density of the IC(Matrix).)  
In each time step we
allow a fraction $\lambda$ of them to leave. In equilibrium the flux into
the sphere (Eq.\ref{eq:Nin}) equals the flux out ($V_s \rho_0 \lambda$),
allowing us to calculate $\lambda$: 
\begin{equation}
\lambda={\frac{2}{3}} {\frac{f_k }{V_s}} \left[ \sqrt{2\pi} \sigma (3 R^2 -
\sigma^2) +2 \pi R^3 \mathrm{\,erfc\,{({\frac{\sqrt{2} R }{\sigma}})}}+ 
\sqrt{2 \pi} \sigma e^{\frac{-2 R^2 }{{\sigma^2}}} (\sigma^2 -R^2) \right]
\label{eq:lambda}
\end{equation}
When $R>>\sigma$, we expect the geometry of the inclusion to be irrelevant.
In this limit, if the random walkers had a \textit{flat} distribution of
steps bounded by $\sigma$, then the flux into the sphere would come from a
thin spherical shell of thickness $\sigma$ and radius $R$. The volume of
this shell is $\sigma A_s$, where $A_s$ is the surface area of the sphere.
The flow in from this shell is balanced by the flow out of the volume, $V_s
\rho \lambda$. It is useful to write this in terms of a new constant, $c_0$,
defined via 
\begin{equation}
c_0 \equiv {\frac{\lambda V }{\sigma A}}
\end{equation}
which is dimensionless and becomes shape independent as $\sigma\to 0$. In
this case 
\begin{equation}
c_0 = f_k \left[ {\frac{1}{\sqrt{2 \pi}}} \left( 1-{\frac{\sigma^2 }{3 R^2}}
\right) + {\frac{R}{3 \sigma}} \mathrm{\,erfc\,{({\frac{ \sqrt{2} R}{\sigma }
})}} \right]
\end{equation}
This quantity is bounded by $f_k/\sqrt{2\pi}$, the result one would get for
an infinite slab. The factor $1/\sqrt{2\pi}$ arises from the fact that
walkers have a gaussian distribution of step sizes, and not a flat one.

\begin{figure}[bt]
\centering
\includegraphics[scale=0.4]{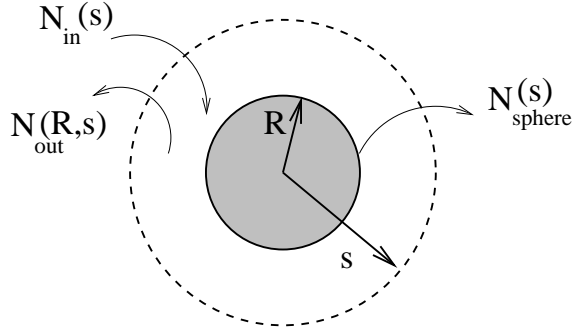}
\caption{{\protect\small An illustration of the three dimensional model for
spheres. The grey material is a sphere of ``infinite'' conductivity with
radius $R$. We wish to calculate the distribution of steps sizes taken by
random walkers leaving the surface of the sphere. We do this by requiring
that in equilibrium the net flux through any surface (e.g. the dashed line
indicating a sphere of radius $s$) must be zero. The number of walkers
hopping in through the dotted surface ($N_{in}(s)$) must be balanced by the
total number of walkers hopping out, both those from the matrix material ($
N_{out}(R,s)$) and those from the surface of the sphere ($N_{\mathrm{sphere}
}(s)$). This condition allows us to calculate the step distribution. }}
\label{fig:sphere}
\end{figure}

Next we have to calculate the distribution of steps for random walkers
leaving the surface of the sphere. As in the one dimensional case of
subsection \ref{subsec:oneD} above, we can calculate the desired result by
balancing fluxes in equilibrium. We draw an imaginary surface of radius $s$
about the spherical inclusion. In equilibrium, the net flux through this
surface must be zero, as illustrated in fig.(\ref{fig:sphere}).

\vspace{25pt} 
\begin{equation}
N_{in}(s)= N_{out}(R,s) + N_{\mathrm{sphere}}(R,s)  \label{eq:sphereBalance}
\end{equation}
The flux in through the sphere of radius $s$, $N_{in}(s)$, is the result eq.(
\ref{eq:Nin}) evaluated for a radius of $s$. The flux outward from the
matrix material is given by the integral: 
\begin{equation}
N_{\mathrm{out}}(R,s) = \int_{R<r<s} d{\vec r} {\frac{\rho_0 f_k }{(2 \pi
)^{3/2} \sigma^3}} \int_{r^{\prime }> s} d\vec r\,^{\prime \frac{-(\vec r -
\vec r\,^{\prime 2 })}{2\sigma^2}}
\end{equation}
We can again evaluate this integral analytically to obtain: 
\begin{eqnarray}
N_{\mathrm{out}}(R,s)= {\frac{2}{3}} f_k \rho_0 \left[ 2 \pi s^3 \mathrm{
\,erfc\,{(\frac{\sqrt{2} s}{\sigma})}} +\sqrt{2 \pi} \sigma (\sigma^2 - s^2)
e^{\frac{-2 s^2}{\sigma^2}} - \sqrt{2 \pi}\sigma (\sigma^2-3 s^2) \right. 
\nonumber \\
\left.+ \sqrt{2 \pi}\sigma (\sigma^2-s^2-R^2-Rs) e^\frac{-(s-R)^2}{2 \sigma^2
} - \sqrt{2 \pi}\sigma (\sigma^2-s^2-R^2+Rs) e^\frac{-(s+R)^2}{2 \sigma^2}
\right.  \nonumber \\
\left. +\pi (s^3-R^3)\mathrm{\,erfc\,{(\frac{s-R}{\sqrt{2}\sigma} )}}- \pi
(s^3+R^3) \mathrm{\,erfc\,{(\frac{s+R}{\sqrt{2}\sigma})}} \right]
\label{eq:NoutRs}
\end{eqnarray}

Finally, we can write an expression for  the flux of random walkers
(originating on the inclusion surface) that hop out through the sphere of
radius $s$: 
\begin{equation}
N_{\mathrm{sphere}}(s)= V_s \rho_0 \lambda \int_s^\infty f(r)\, dr
\label{eq:Nsphere}
\end{equation}
where $f(r)\, dr$ gives the fraction of walkers that jump radially outward
to a distance between $r$ and $r+dr$ from the center of the inclusion.

Eqns.(\ref{eq:Nin}), (\ref{eq:NoutRs}) and (\ref{eq:Nsphere}) give us enough
information to calculate the step distribution function $f(r)$. However in
computer applications we do not actually use $f(r)$. Rather algorithms
typically generates a random number, $p$, in a flat distribution $0<p<1$,
and use that to select a random step $\delta(p)$ from the center of the
sphere, $\delta>R$. We can do this by first calculating the integral of $f(r)
$: 
\begin{equation}
P(\delta) \equiv \int_R^{\delta} f(r^{\prime }) \, dr^{\prime }
\label{eq:integdist}
\end{equation}
Note that $P(R) = 0$ and $\lim_{\delta\to \infty} P(\delta) = 1$. We then
must invert this functional relationship to get $\delta(P)$, which gives us
the step generating function we desire. We note that from 
eq.(\ref{eq:Nsphere}) and (\ref{eq:integdist}) we have: 
\begin{equation}
N_{\mathrm{sphere}}(R,s) = V_s\, \rho_0 \lambda \left[ 1-P(s) \right]
\end{equation}
Equating this via eq.(\ref{eq:sphereBalance}) and dropping exponentially
small terms we have

\begin{eqnarray} 
P(s) & = & 1-  \nonumber \\ 
& & {} \left[ \pi(R^3-s^3)
\mathrm{\,erfc\,}{\left({\frac{{s-R}}{{\sqrt{2} \sigma}}}\right)}
- \sqrt{2 \pi}\sigma (\sigma^2-s^2-R^2-Rs) e^{\frac{{ -(s-R)^2}}{{2
\sigma^2}}} \right.  \nonumber \\ & & \frac{ {} + \left. \sqrt{2
\pi}\sigma (\sigma^2-s^2-R^2+Rs) e^\frac{ -(s+R)^2}{2 \sigma^2}
+\pi (s^3+R^3)\mathrm{\,erfc\,}{\left(\frac{s+R}{\sqrt{2
}\sigma}\right)}\right] }{{\displaystyle {\sqrt{2 \pi} \sigma (3R^2
-\sigma^2) +\sqrt{2 \pi} \sigma (\sigma^2-R^2)e^{\frac{-2 R^2}{\sigma^2}}
+ 2\pi R^3 \mathrm{\,erfc\,}\left(\frac{\sqrt{2}R}{\sigma}\right)}}}
\label{eq:Prob3D} 
\end{eqnarray}

This result has the desired behavior at the limits, $P(R)=0$ and $\lim_{s\to
\infty} P(s)=1$. This function is not analytically invertible; in
implementation it is evaluated on a mesh and the inverse is calculated via
interpolation.

\subsection{Numerical Results in 3D \label{subsec:NumResults3D}}

We implemented a random walk algorithm in three dimensions similar to that
of section \ref{sec:1D} above. In three dimensions we applied periodic
boundary conditions in the $y$ and $z$ directions. A temperature profile in
the $x$ direction was obtained simply by binning all walkers in a given
range of $x$ for all $y$ and $z$; such slices would cross inclusions as well
as matrix material. Walkers that were labelled as inside a given inclusion
were assigned a random position inside the inclusion for the purpose of
doing this averaging. The simulation volume was $10\times 10 \times 10$, and
the random walk in the matrix was described by a Gaussian distribution with
a rms value of 0.10 in these units. The transition probability $f_{m,IC}$
was fixed at 1.0.

\begin{figure}[bt]
\centering
\vbox{\includegraphics[width=3.25in]{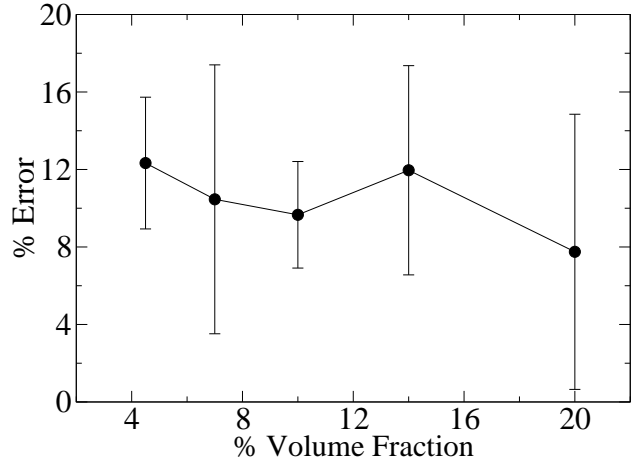}}
\caption{{\protect\small  Plot of the percent error in the thermal
conductivity as a function of the volume fraction for fixed surface area.
The percent error is defined as the ratio of the difference of thermal
conductivities measured using the Gaussian step distribution and that of eq.
(\protect\ref{eq:Prob3D}) divided by the former. Note that the error varies
only slightly with volume fraction. }}
\label{fig:evn3D}
\end{figure}

In fig.(\ref{fig:evn3D}) the percent error (defined as the ratio of the
difference of thermal conductivities measured using the Gaussian step
distribution and that of eq.(\ref{eq:Prob3D}), divided by the former) is
plotted as a function of the volume fraction of infinite conductivity
inclusions, for a fixed surface area for the inclusions. (If there were
only a single spherical
inclusion, it would have had a volume fraction of 5\%.)\cite{fudge} As the number of inclusions at fixed surface area increases, their
total volume decreases as $N^{-3/2}$. (For example, the largest volume
fraction, 0.20, corresponds to $100$ spheres of radius $0.9772$). The
results at several values of $N$ were calculated for five random
configurations and the average and standard deviation are plotted. The
effect of using a simplified step distribution is \textit{larger} in three
dimensions, and can affect the results by up to 18\%.

\begin{figure}[bt]
\centering
\vbox{\includegraphics[width=3.25in]{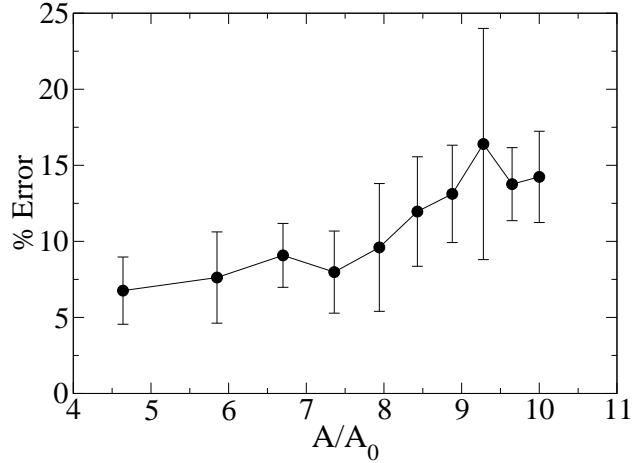}}
\caption{{\protect\small Plot of the percent error in the thermal
conductivity as a function of the surface area of the spherical inclusions
at a fixed volume fraction of 5\%. The surface area is measured in terms of $
A_0$, the surface area of a sphere with 5\% of the total volume. The percent
error is defined as the ratio of the difference of thermal conductivities
measured using the Gaussian step distribution and that of 
eq.(\protect\ref{eq:Prob3D}). divided by the former. }}
\label{fig:eva3D}
\end{figure}

The percent error was also calculated as a function of the surface area for
fixed volume fraction and plotted in fig.(\ref{fig:eva3D}). The volume
fraction was fixed at 5\%, and the surface area increases with $N$ as $
N^{2/3}$. Five simulations were run for $N=100, 200, \dots 1000$ and the
average and standard deviation were plotted as a function of the surface
area relative the minimum surface area, $A_0$, the area of a sphere that is
5\% of the volume. Again, the effect of using the wrong simulation algorithm
is shown to be substantial.

\section{Conclusions and Future work \label{sec:Conclusions}}

Transport in composites with a large disparity in conductivities is
important to a large number of systems. In this paper we have demonstrated
an efficient and physically sound algorithm for calculating effective
conductivities of composites with large contrasts in conductivity. We have
shown that the errors introduced are small but measurable in one dimension,
and moderately significant in 3D.

The spherical inclusion case is the simplest 3D problem, but not the most
relevant to many systems. Carbon nanotubes might be approximated as
cylinders, to lowest order. However, in that case the 1D integrals of
section(\ref{subsec:Sphere}) become more complicated and handling the
endcaps of the cylinders becomes problematic. A simple approach might be to
simply ignore transport through the endcaps, or treat a nanotube as an
extremely prolate spheroid so that diffusion from the inclusion can again be
treated as a one dimensional walk normal to the surface. These
approximations are the subject of current research.

% -----------------------------------------------------------------------------

\begin{acknowledgments}
This project was supported in part by the US National Science Foundation
under Grant~\mbox{MRSEC DMR-0080054}, and \mbox{EPS-9720651}, 
and~\mbox{PHY--0071031}. Dimitrios Papavassiliou acknowledges support from the
DoE-funded Carbon Nanotubes Technology Center - (CANTEC, Award Register\#:
ER64239 0012293).
\end{acknowledgments}

% -----------------------------------------------------------------------------

%\begin{appendix}
%\end{appendix}
% -----------------------------------------------------------------------------

%\clearpage
%\bibliographystyle{apsrev}
%\bibliography{/biblios/physjabb,/biblios/NO,/biblios/books,/biblios/misc,newrefs,BEC07172003,/biblios/Ultracold}

\begin{thebibliography}{99}
\bibitem{ESChoi} E.S.Choi, J.S.Brooks, and D.L.Eaton, J.Appl.Phys. \textbf{94
} 6034(2003).

\bibitem{JChe} J.Che,T.Cagin, and W.A.Goddard III,Nanotechnology(2003).

\bibitem{JHone} J.Hone,B.Batlogg, Z.Benes, A.T.Jonson, and J.E.Fischer,
Science \textbf{289}, 1730(2000).

\bibitem{SBerber} S.Berber, Y.K.Kwon, and D.Tomanek, Phys. Rev. Lett. 
\textbf{84},4613(2000).

\bibitem{RBBird} R.~B.~Bird, W.~S.~Stewart, and E.~N.~Lightfoot, \textit{
Transport Phenomena}, 2nd ed. (Wiley, New York, 2002),pp.282,376 and 397.

\bibitem{elect1} %The Electrical Conductivity of Composite Media
E.~H.~Kerner, 1956 Proc. Phys. Soc. B 69 802

\bibitem{elect2}  Lawrence E. Nielsen,
Ind. Eng. Chem. Fundamen., 1974, 13 (1), 17.

\bibitem{Toma1} M.~M.~Tomadakis and S.~V.~Sotirchos, J.~Chem.~Phys. \textbf{98}, 616 (1992).

\bibitem{Toma2} M.~M.~Tomadakis and S.~V.~Sotirchos, J.~Chem.~Phys. \textbf{104}, 6893 (1996).

\bibitem{cnt1} M.~H.~Duong, D.~V.~Papavassiliou, K.~J.~Mullen and L.~L.~Lee,
Appl.~Phys.~Lett. \textbf{87}, 013101(2005).

\bibitem{cnt2} M.~H.~Duong, D.~V.~Papavassiliou, K.~J.~Mullen and S.~Maruyama,
Nanotechnology \textbf{19}, 065702 (2008).

\bibitem{cnt3} H.~M. ~Duong, D.~ V. ~Papavassiliou, K.~ J.~Mullen, B.~ L.~
Wardle, S.~Maruyama,  Inter. J. of Heat and Mass Transfer, 52
(2009) 5591-5597.

\bibitem{Kap} P.~L.~Kapitza, J. Phys. (USSR) \textbf{4},181 (1941).

\bibitem{Shen} S.~Shenogin, L.~Xue, 
R.~Ozisk, P.~Keblinski, and D.~G.~Cahill,
J.~Appl.~Phys. \textbf{95}, 8136(2003).

\bibitem{JLB} J.~L.~Barrat and F. Chiaruttini, Mol.Phys. \textbf{101}
,1605(2003).

\bibitem{CJTwu} C.~J.~Twu and J.~R.~Ho, Phys.~Rev.~B \textbf{67}, 205422 (2003).

\bibitem{SchwartzPohl}  E.~T.~Swartz, and R.~O.~Pohl,  Rev. Mod. Phys. 1989,
{\bf 61}  (3), 605. 

\bibitem{probability} The function $\tilde f(u)$ gives the probability of a
step of size $u$. In a Monte Carlo algorithm, one must calculate the inverse
of the function $g(p)=\int_0^p \tilde f(u) \,du$, and use it to convert a
flat distribution of random numbers $0<p<1$ into a the correct random step
size.

\bibitem{fudge} The volume used in calculating the volume fraction is
slightly smaller that the total simulation volume, because the spheres were
excluded from the two thin layers where the walkers were injected.
\end{thebibliography}

\clearpage
% -----------------------------------------------------------------------------

%\newpage
%\printtables % remove when remove endfloats option
%\newpage
%\printfigures % remove when remove endfloats option

\end{document}